\documentclass[letterpaper, 10 pt, conference]{IEEEtran}  

\IEEEoverridecommandlockouts                              

\usepackage{xcolor,soul}

\usepackage{amsmath} 
\usepackage[caption=false]{subfig}
\usepackage{graphicx}
\usepackage{lipsum}

\title{\LARGE \bf
Ground Roll Suppression using Convolutional Neural Networks
}

\author{Dario A. B. Oliveira$^{1}$, Daniil G. Semin$^{2}$ and Semen Zaytsev$^{2}$
\thanks{Dario A. B. Oliveira is with IBM Research,
        Av. Pasteur 138/146, 22290-240, Botafogo, Rio de Janeiro, Brazil
        {\tt\small dariobo at br.ibm.com}}
\thanks{$^{2}$Daniil G. Semin and Semen Zaytsev are with Gazprom Neft {\tt\small \{semin.dg, zaytsev.sale\}  at gazpromneft-ntc.ru}}

}

\begin{document}


\maketitle
\thispagestyle{empty}
\pagestyle{empty}

\begin{abstract}
Seismic data processing plays a major role in seismic exploration as it conditions much of the seismic interpretation performance. In this context, generating reliable post-stack seismic data depends also on disposing of an efficient pre-stack noise attenuation tool. Here we tackle ground roll noise, one of the most challenging and common noises observed in pre-stack seismic data. Since ground roll is characterized by relative low frequencies and high amplitudes, most commonly used approaches for its suppression are based on frequency-amplitude filters for ground roll characteristic bands. However, when signal and noise share the same frequency ranges, these methods usually deliver also signal suppression or residual noise. In this paper we take advantage of the highly non-linear features of convolutional neural networks, and propose to use different architectures to detect ground roll in shot gathers and ultimately to suppress them using conditional generative adversarial networks. Additionally, we propose metrics to evaluate ground roll suppression, and report strong results compared to expert filtering. Finally, we discuss generalization of trained models for similar and different geologies to better understand the feasibility of our proposal in real applications.

\end{abstract}

\section{Introduction}

Seismic data acquisition is a complex and expensive process consisting of a set of geo-located geophones that record vibrations from the earth followed by a controlled stimulus, such as an explosion. In this process, geophones record not only reflected seismic signals, which provide information from inside the earth, but also undesired random and coherent noises. One of the most common and challenging coherent noises in land seismic imaging is ground roll, which is characterized by its relative high amplitudes and low frequencies in comparison with the usual seismic signal. Ground roll noise is usually dominated by Rayleigh waves but often present components of Love waves, reverberated refractions, and waves scattered by near-surface heterogeneities.  

The simplest approach to cope with ground roll noise is applying a low-cut frequency filter, but since ground roll shares some of the low frequencies observed in seismic signals, it severely impacts the quality of subsequent steps of seismic processing, and therefore final stack overall quality. Considering these ordinary low frequencies, most of commonly used methods needs to choose between suppressing aggressively ground roll noise (and attenuate also signal), or protecting the seismic signal (and leave part of ground roll noise present). Domain transformation is very common, with popular techniques being widely used like f-k filtering \cite{yilmaz}, Radon transform \cite{henley} or frequency range noise attenuation using LIFT technology \cite{Jason}.




Deep learning techniques became recently popular in the geoscience community, and works to process seismic data have been proposed in last few years. While \cite{oliveira} used generative models for improving resolution of seismic data, \cite{na1} designed a neural network for filtering ground roll from two shots in advance and then applying it to all shots of the field land shot gather, and \cite{na2} explore different architectures to cope with common seismic noises, presenting limitations and results for synthetic and field data.

In this paper, we introduce the combined use of detection, segmentation and adversarial networks to tackle ground roll in shot gathers. We implement a two-step pipeline: first we detect the ground roll region using a combination of xCeption to detect roughly the affected area and a UNet segmentation network to segment ground roll in individual seismic traces; then we apply filtering only in the affected area, using conditional generative adversarial networks (cGANs) that allow noise filtering while restoring characteristic signal amplitudes and frequencies. This simple yet efficient pipeline allows us to fully preserve signal where no ground roll is observed and at the same time be as aggressive as possible for ground roll suppression in the affected area.  

Our work is organized as follows: we present our methodology in section \ref{sec:method}, providing details on the proposed detection and filtering pipelines. Then, we delineate our experimental design and discuss our results in section \ref{sec:experiments}, and finally draw conclusions in section \ref{sec:conclusions}.  


\section{Methodology}
\label{sec:method}

Our methodology consists of two major steps: first, we tackle ground roll as a noise detection problem 
using detection and segmentation neural networks; then we use conditional generative adversarial networks to filter ground roll noise in affected regions, letting non-affected regions untouched. Each of these steps are detailed in the following.


\subsection{Ground Roll Detection}

The unsupervised detection pipeline herein proposed consists of 3 steps: seismic data pre-processing, rough ground roll detection and fine ground roll segmentation.

\subsubsection{Pre-processing}

The large difference in amplitude ranges observed between ground roll noise and signal leads to very low contrast of signal areas, which is very problematic for training convolutional neural networks, since their optimization relies on the backpropagation of gradient error values. To cope with that, we apply a standard histogram equalization that allows the values observed in the data to have a uniform histogram distribution, and apply the same transformation to all gathers.

\subsubsection{Rough ground roll detection}

We split ground roll detection in two sequential steps: first we use a detection network to roughly detect the affected area, then we use the resulting mask as training data for a segmentation network that segments ground roll at each individual seismic trace.

\begin{figure}[!ht]
\centering
\includegraphics[width=0.8\linewidth]{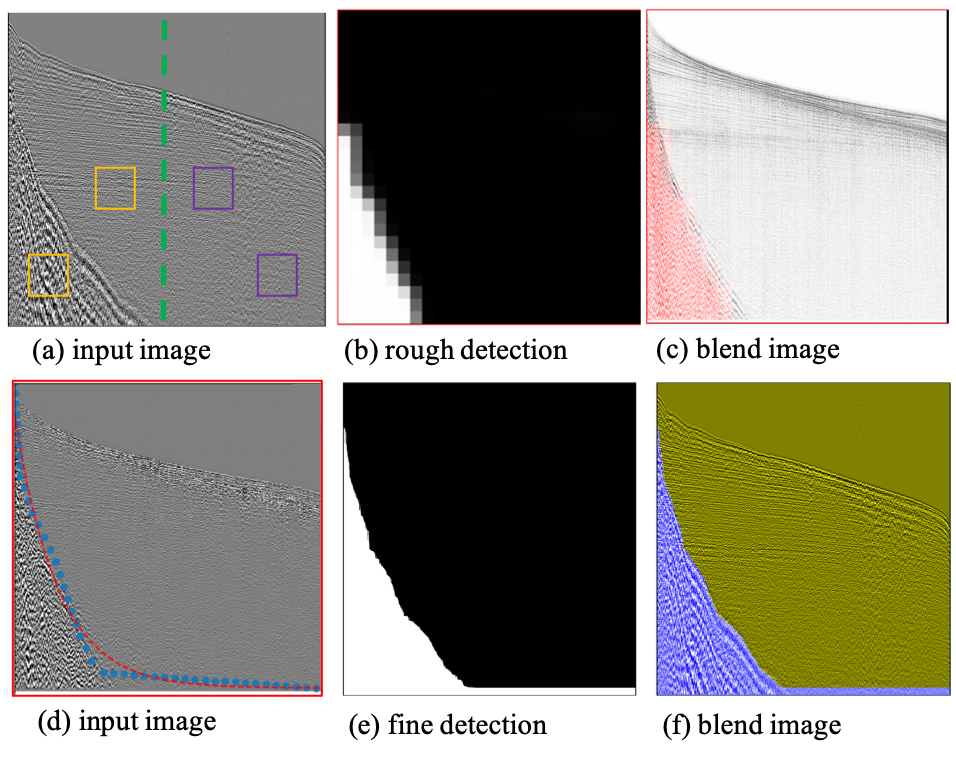}
\caption{Rough detection of ground roll affected area. (a) depicts the input data and some landmarks: green dashed line indicates half of data used for training, where yellow tiles are noise-affected samples and purple tiles are noise-free samples. (b) presents the output of xCeption network for likelihood of being a noise-affected tile. (c) presents a blend of (a) and (b). (d) presents the input data and red dashed line is the final rough ground roll detected area after the logarithmic function fit used for training. (e) presents the outcome mask of UNet pos-processed data, (f) presents a blend of (d) and (e).}
\label{fig:rough_det}
\end{figure}

To implement rough detection, we used a 2D xCeption detection network \cite{xception}, that learns to identify if a given tile of the data is affected by ground roll noise or not. We use the fact that for the studied area ground roll happens mainly in the first half of a shot gather with offset ordering and assume that the second half would contain basically noise-free information, as shown in Figure \ref{fig:rough_det}(a). With this heuristic - that could be easily manually defined - we developed an experimental design that samples noise-free tiles (purple coloured) from the unaffected half area and noisy tiles (yellow coloured) from the affected half area of seismic data, as depicted in Figure \ref{fig:rough_det}(a). Since this assumption is roughly true, at convergence, the network should learn to differentiate noisy and noise-free tiles, even if part of the training set is incorrectly labeled. 


To cope with geometry inconsistent masks caused by detection of tiles partially affected by ground roll, we fit a logarithmic function to the detection image and create a more robust mask for training the segmentation network. In figure \ref{fig:rough_det}(d), we observe the points used to fit the function in blue, and the logarithmic function fit as red dashed line. 


\subsubsection{Fine ground roll detection}


With the rough affected area detection, we train a UNet semantic segmentation network \cite{Ronneberger2015} - modified to cope with 1D data - using the training set created from the logarithmic function masks: samples are 1D arrays, 0-valued in ground roll region in the seismic trace, and 1-valued in non-affected region. Again, we expect some annotation imprecision because the logarithmic function does not perfectly match the ground roll affected area. Still, at convergence, the network should be able to learn what differs ground roll noise from signal and correctly segment the traces. With this pipeline, we are able to create masks of ground roll in seismic gathers very consistent with the visually expected, as shown in Figure \ref{fig:rough_det}(d) and (e).



\subsection{Ground Roll Filtering}

After detecting the ground roll region, a filtering method is applied to remove the noise in the affected region, while preserving the original information outside the mask. Our supervised filtering pipeline consists of 4 steps, detailed in the following: pre-processing, tiles sampling, cGANs filtering and pos-processing.

\subsubsection{Pre-processing}

As presented in detection section, we applied a standard histogram equalization to all gathers.

\subsubsection{Tiles sampling}
To sample tiles for creating the training set in the supervised training, we simply select a number of points inside the ground roll affected region and extract paired samples: one from the original ground roll gather and the other from the manually filtered reference gather. This way we create a list of paired tile samples, which are used to train a network that will learn to map noisy tiles into the corresponding clean tiles.

\subsubsection{cGANs filtering}


For removing ground roll we use an image translation approach, introduced by Goodfellow et al.~\cite{Goodfellow2014} in 2014: generative adversarial networks (GANs). Adversarial training involves two convolutional networks: the generator ($G$), responsible for synthesizing images ($y$), and the discriminator ($D$) that learns if a given image is real or synthesized. The main advantage of these networks is that they dedicate a whole network to model the loss function (the discriminator), which allow loss to be much more complex than the usual mean square error functions or alike. This allows the combined network to find better solutions in the solution space by forcing the generator to create very realistic outputs.

More specifically, we used a type of GANs that targets paired image translation: conditional GANs (cGANs), introduced by \cite{Isola2016}. These networks use both the input and the paired target image to train the networks: while the generator tries to map the input into the target, the discriminator receives both input and target or generated image to learn which pair of images is real or fake. After the training procedure, it is expected that the generator network is able to generate very realistic target images from inputs. In our training schema, we map ground roll affected tiles into reference clean tiles.

Formally, in cGANs, the generator learns to create images from a domain ($y$) using images from another domain ($x$), and the discriminator learns to identify if a given image from $y$ is real or synthesized, given a pair of images from $x$ and $y$. The loss function for cGANs is presented in Equation~\ref{equ:cgansobjective}. Considering a training set distribution $p_{data}(x)$, the trained generator learns a similar distribution $p_{model}(w)$ such that the discriminator is actually not able to detect if a given sample came from $p_{data}(x)$ or $p_{model}(w)$.
The optimization formulation for finding the optimal mapping function $G^{*}$ is usually defined as
     \begin{equation} \label{equ:Lgan}
        G^{*} = \mathrm{arg}\; \mathrm{min}_{G}\;\mathrm{max}_{D}\; \mathcal{L}_{GAN}(G,D)
    \end{equation}
    where $L_{GAN}(G,D)$ is the cGANs objective function, 
    \begin{equation} \label{equ:cgansobjective}
        \begin{split}
        \mathcal{L}_{cGAN}(G,D) &= E_{x,y\sim p_{data}(x,y)}[log D(x,y)]+  \\
                      &\; E_{x\sim p_{(x)},z\sim p_{(z)}}[log(1 - D(x,G(x,z))]
        \end{split}
    \end{equation}
    
The solution of Equation~\ref{equ:cgansobjective} is achieved using an alternate training schema: the generator training uses the loss obtained using the last trained discriminator, while the discriminator computes losses using images synthesized by the last trained generator. 

We used the Pix2Pix cGAN architecture, which implements a generator with an encoder-decoder architecture \cite{Ronneberger2015} and a ``PatchGAN''discriminator that uses patches to incorporate local information to the loss function, and can be fully understood in the original paper ~\cite{Isola2016}.




\subsubsection{Pos-processing}


Our pos-processing involves amplitude correction and replacing the ground roll detected area with the filtered outcome from our approach. 
To correct amplitudes in the filtered area, we perform a masked histogram equalization, by computing the signal region cumulative density probability function and resampling amplitude pixel values at the noise region using the same density probability and values. This way, amplitude values distribution observed in the filtered region will be very similar to the one observed in signal region. 


\section{Experiments}
\label{sec:experiments}

\subsection{Training and Test Data}

In our experiments, we used 3 real 3D seismic datasets from different Western Siberia oilfields: 2 from the HMAO (dataset 1 and 2) and 1 from the YaNAO (dataset 3) regions. Since Western Siberia is a vast territory, the near surface conditions as well as geological structure of the HMAO and the YaNAO regions differ quite strongly, and this causes a difference in ground-roll characteristics for these regions. In addition, for YaNAO ground-roll is observed throughout all gather up to the far offsets, while in HMAO data, only at the near offsets.

All datasets used explosive type of source and were recorded with 6s trace length of 2ms sample interval using the orthogonal acquisition pattern with Shot/Receiver 50m points dense while line spacing at 300m. Number of channels, receiver types, maximum offsets, fold differ from field to field, as observed in table \ref{fig:tab-data}.

\begin{table}[ht]
\centering
\caption{Datasets characteristics: regions, number of channels, elevation ranges, ground-roll frequencies, maximum offsets of ground roll, number of folds.}
\begin{tabular}{|c|c|c|c|c|c|}
\hline
Dataset & Channels & Elevation & GR Freq. & Max offsets & Fold \\ 
\hline
1-HMAO & 3400 & 27-44m & 5-20Hz & 2500/5500  & 140 \\ 
2-HMAO & 800 & 25-80m & 5-25Hz & 1400/3000 & 60  \\ 
3-YaNAO & 2100 & 100-220m & 6-24Hz & 4000/4000 & 100  \\ 
\hline
\end{tabular}
\label{fig:tab-data}
\end{table}


\subsection{Training scheme}



All training rounds used the same configuration: 100 epochs (a safe margin, since even fewer epochs already lead to satisfying results), each epoch considering 50{,}000 samples taken at random from 5 different shot gathers, using Adam optimizer with initial learning rate of $0.0002$ and momentum of $0.5$. For filtering networks, generator and discriminator loss functions are detailed in~\cite{Isola2016}, and the parameters of generator loss function were set to: GAN weight $=1$ and L1 weight $=100$. 

\subsection{Evaluation metrics}
\label{sec:metrics}

Evaluation of automatic methods to compare with human outcome is not simple because some metrics not necessarily implement the expert expected visual outcome. Here, we propose three different metrics, two of which - power spectrum analysis and amplitude analysis - are often used to evaluate ground roll suppression in the industry, and the other - trace correlation coefficient - evaluates directly the agreement with human outcome, detailed below:
\begin{itemize}
    \item Power spectrum mean distance: this metric is computed considering the distance between spectra of signal and noise regions. Ideally, both spectra should be very similar, and distance would be small. 
    \item Amplitude range analysis: differences of amplitude ranges between noise and signal regions should not happen, and ideally should be zero.
    \item Correlation coefficient with reference: trace correlation between ML filtered and reference data should be close to 1, if the reference is perfect. 
\end{itemize}

Considering the power spectrum, for each reference (raw gather, reference filtered data, and result data) we compute two power spectrum density (periodogram): one for signal region and the other for noise region considering a frequency range of 0-60Hz. Each periodogram is then represented by a discrete set of power values $P_i$ for each frequency $i$. $P(S)_i$ is the periodogram for signal area and $P(N)_i$ for noise area. Power spectrum density average distance for each reference is:
    \begin{equation} \label{equ:pwr}
        P_D = \frac{\sum\limits_{i=0}^50 abs(P(S)_i - P(N)_i)}{n}
    \end{equation}
    where the amplitude score is given by, 
    \begin{equation} \label{equ:pwr_score}
        Q_p = \frac{P_D(original)-P_D(result)}{P_D(original)-P_D(expert)} \times 100\%
    \end{equation}
    
Power spectrum score $Q_p$ for a given gather is computed using as reference the density average distance observed in the reference data. Values above $100\%$ can be seen as results where the average distance is lower than the observed in the reference, meaning they have frequency spectrum closer to the signal area than the observed in reference data.

Considering the amplitude analysis, for each reference (raw gather, reference filtered data, and result data) we compute the frequency density (histogram) of amplitude values for each window (signal vs noise). Each histogram is then represented by a discrete set of amplitude density values $H_i$ for each amplitude bin value $i$. $H(S)_i$ is the histogram for signal area and $H(N)_i$ for noise area. Then, the histogram average distance for each reference is given by:
    \begin{equation} \label{equ:amp}
        H_D = \frac{\sum\limits_{i=0}^n abs(H(S)_i - H(N)_i)}{n}
    \end{equation}
    where the amplitude score is given by, 
    \begin{equation} \label{equ:amp_score}
        Q_a = \frac{H_D(original)-H_D(result)}{H_D(original)-H_D(expert)} \times 100\%
    \end{equation}
    
Amplitude score $Q_a$ for a given gather is also computed using as reference histograms distance observed in the reference data with respect to distances observed in original unfiltered data. Again, values above $100\%$ can be interpreted as results that have amplitude closer to signal area than the observed in reference.

Considering the correlation coefficient, we compute the average mean trace correlation coefficient (CC) between reference and result for each seismic trace in the gather. Then, for each gather we compute the average value for all traces, and this is the final CC value for a given gather. The score for correlation is set as $Q_c = CC \times 100 \%$
 

\subsection{Experimental results: individual gather evaluation}

We begin our experimental results section presenting evaluation of an individual gather. Since it has a didactic objective, we used an example of a similar geology but different area (dataset 2 from Table \ref{fig:results-tab}) from the data used for training the models (dataset 1 from Table \ref{fig:results-tab}). Each shot gather is evaluated considering the metrics presented in section \ref{sec:metrics}. 

\begin{figure}
\centering
\includegraphics[width=0.8\linewidth]{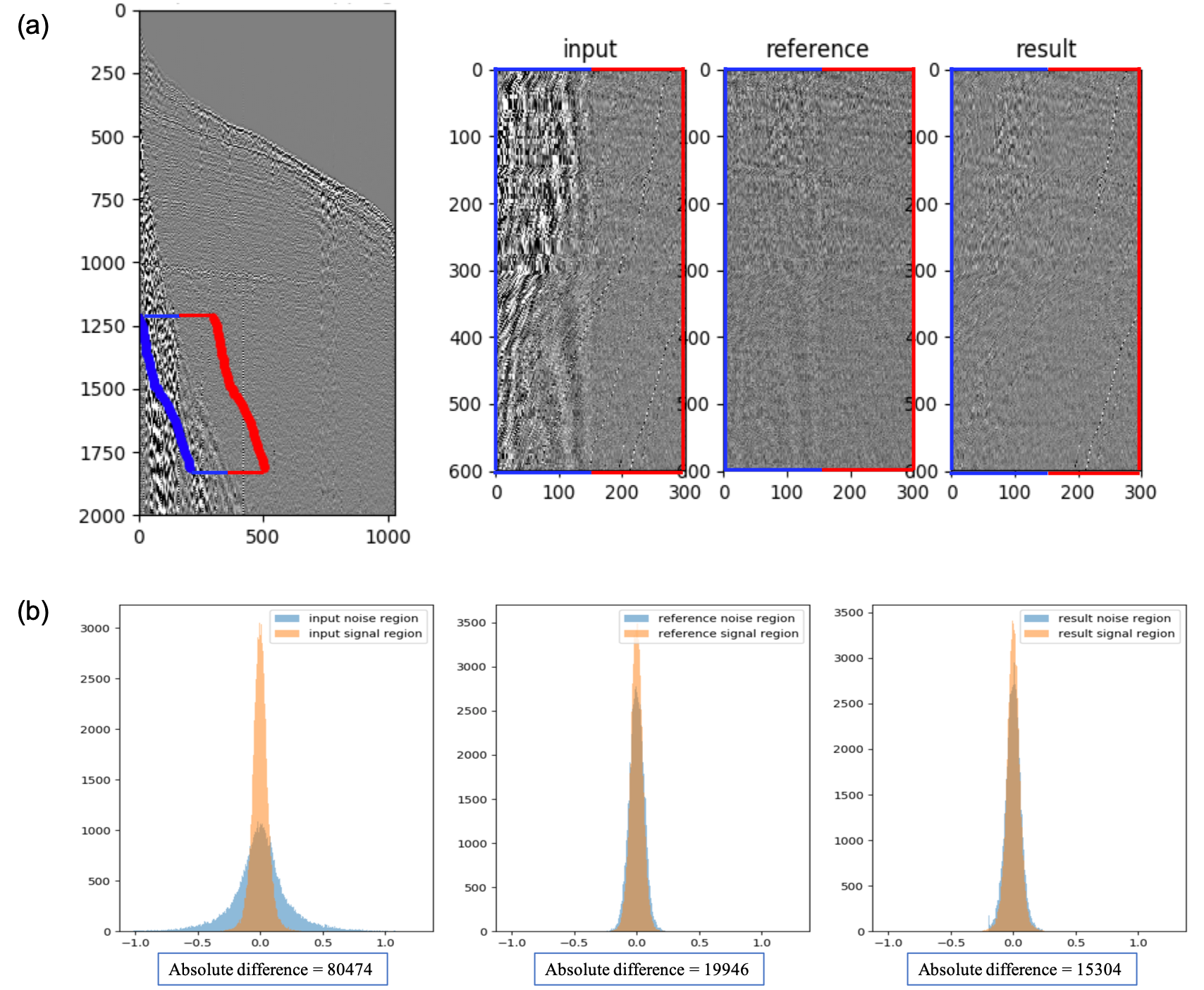}
\caption{Amplitude analysis: in (a) we observe the window analysis, and in (b) the amplitude histogram of signal and noise/filtered regions for original, result and reference data.}
\label{fig:amplitude}
\end{figure}

\begin{figure}
\centering
\includegraphics[width=0.8\linewidth]{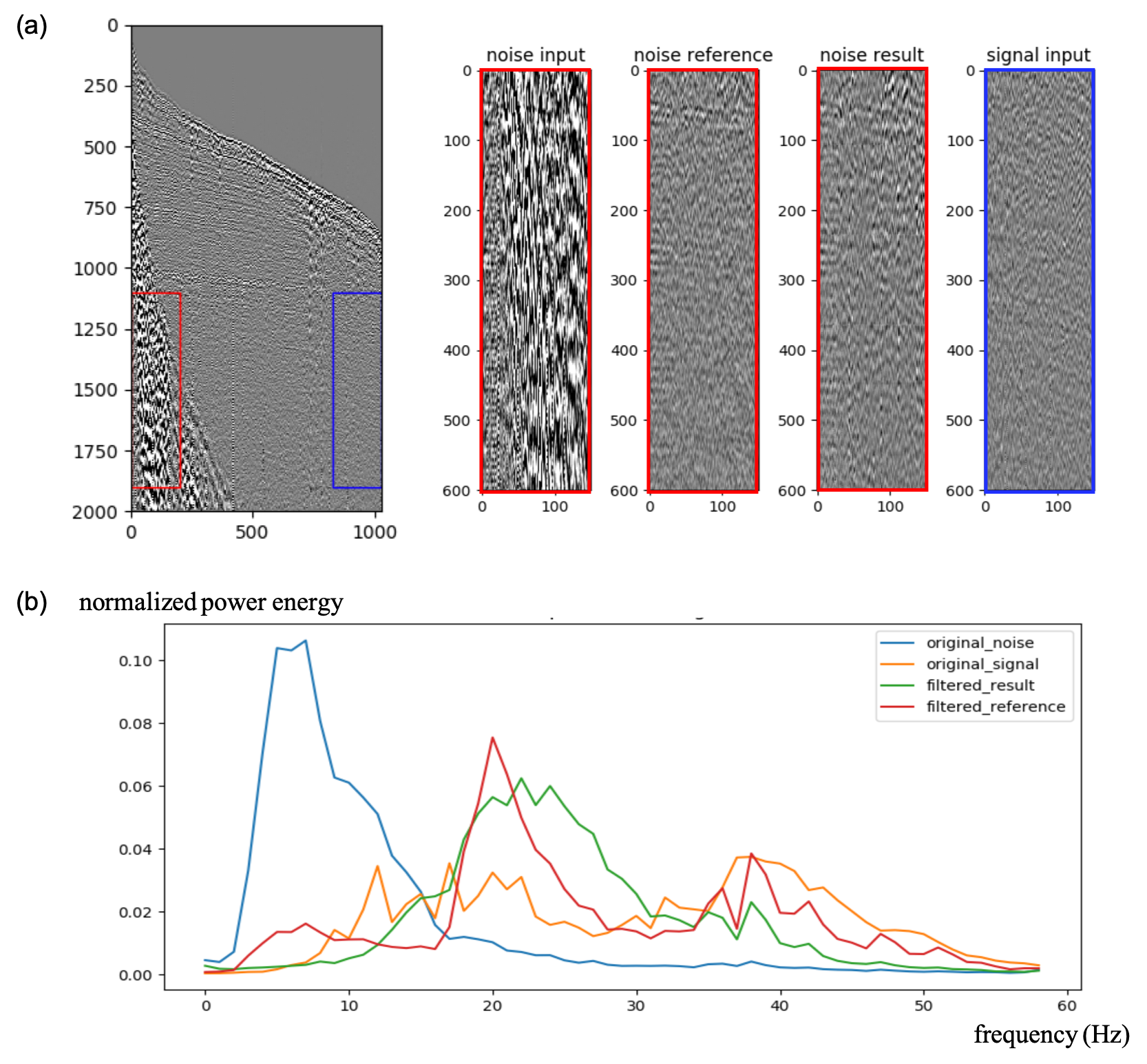}
\caption{Amplitude analysis: in (a) we observe the window analysis, and in (b) the power spectrum of signal and noise/filtered regions for original, result and reference data.}
\label{fig:power}
\end{figure}

\begin{figure}
\centering
\includegraphics[width=0.75\linewidth]{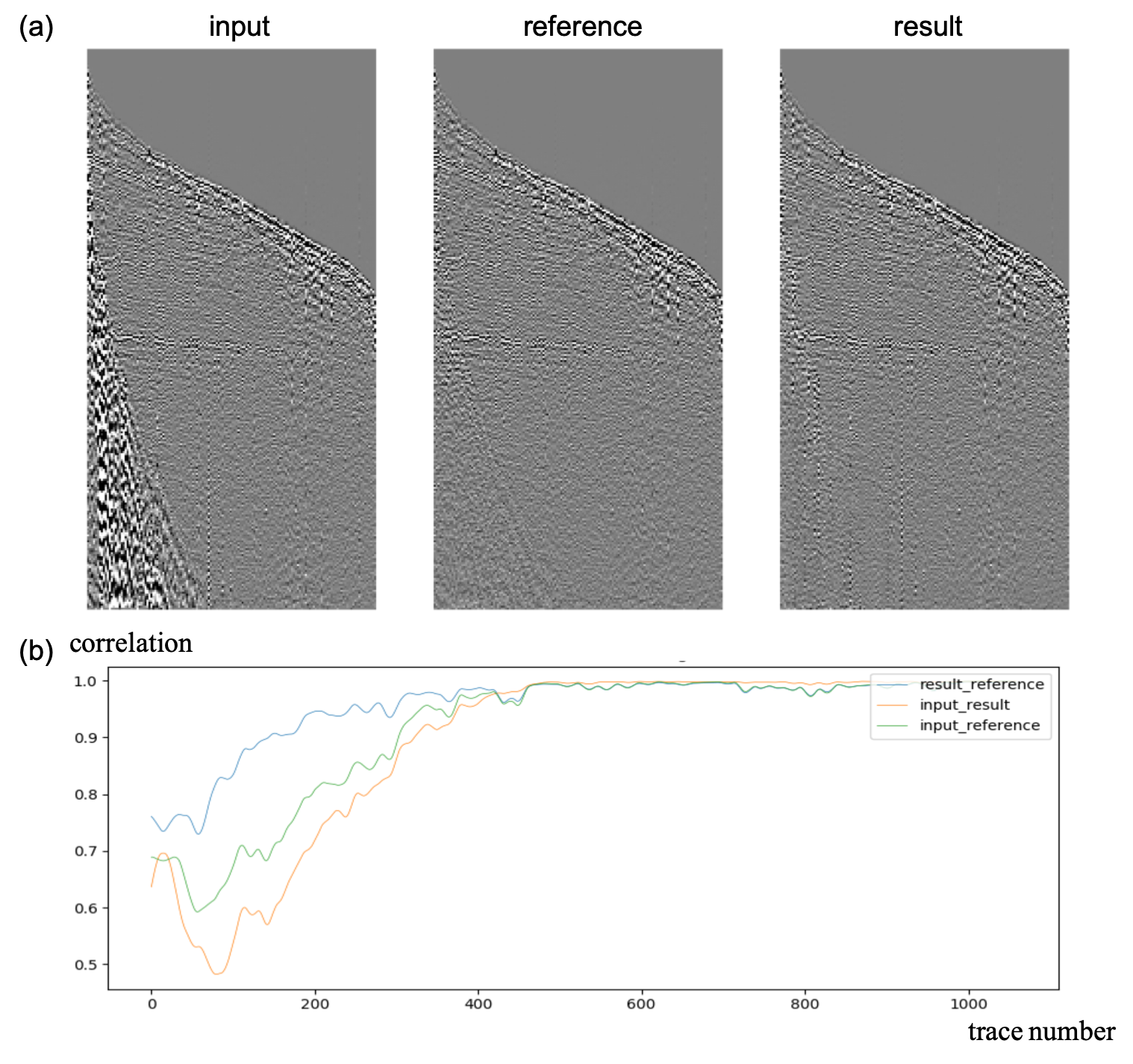}
\caption{Correlation coefficient analysis: in (a) we observe the input, reference and filtered gathers, and in (b) the trace correlation coefficient chart, where each $x-axis$ value is a trace index in offset ordering, and the corresponding correlation coefficient observed in $y-axis$}
\label{fig:correlation}
\end{figure}

One important aspect of such evaluation is the windows considered for computing the metrics. For amplitude analysis we considered a window embracing the detection mask border, because the aspect to be evaluated here is a desired smooth amplitude transition from filtered to original signal data, as Figure \ref{fig:amplitude} depicts. For power spectrum analysis we considered two distant windows that undoubtedly hold filtered and signal regions, presented in Figure \ref{fig:power}. Finally, the correlation coefficient is applied to every seismic trace in the gather, as shown in Figure \ref{fig:correlation}.

Considering the amplitude, one can visually inspect in the three windows from Figure \ref{fig:amplitude}(a), that our result is very similar to the reference. As explained before, we numerically evaluate the result by comparing amplitude histograms of original, reference and result data. It is also possible to inspect that the histogram in our results is the most similar considering signal and noise (filtered) regions. The average amplitude score for this individual gather, computed as presented in section \ref{sec:metrics} is $107.67\%$. 

Considering the power spectrum, we show windows of noise, reference, filtered and signal in Figure \ref{fig:power}(a). One can visually inspect that our results are quite similar to the reference and signal windows, very different from the original noise. We also present in Figure \ref{fig:power}(b) the plots for power spectrum for each window, where we can clearly see a peak in lower frequencies for original noise, and similar curves for reference, result and original signal windows. It is interesting to notice that our method was able to remove lower frequencies better than the human counterpart. The average power spectrum score for this individual gather, computed as presented in section \ref{sec:metrics} is $75.54\%$. 

Considering the correlation coefficient, present the resulting gather data in Figure \ref{fig:correlation}(a), and the corresponding coefficients chart in \ref{fig:correlation}(b). One can assess that correlation is lower in the first traces, since we are using shot gathers, and ground roll is more severe in this area. Still, our method is able to deliver correlation always above 0.7, which indicates a high quality result. We also note that our result correlates less with noise than the reference, and correlates more with original signal in noise-free areas, which seems to be positive. The average correlation score for individual gather shown in Figure \ref{fig:correlation}, computed as presented in section \ref{sec:metrics} is $95.7\%$. 

\subsection{Experimental results: survey evaluation}

Since we used only a few gathers to train the models, it is important to evaluate how the trained models behave in all the other gathers in the full survey. One would expect models to fairly generalize in similar geology and conditions, which is more likely to happen in the same survey.

The overall average scores for the survey used for training the model (dataset 1) are: amplitude score of $104.51\%$, power spectrum score is $118.99\%$, and trace correlation score is $94.86$, as depicted in Table \ref{fig:results-tab}. It is important to mention that this results above 100\% does not necessarily mean that our results are better than human filtering in general, but only that our pipeline was able to deliver patterns of power spectrum and amplitude closer to the ones observed in signal regions. Results from power spectrum and amplitude also indicate that differences observed in the trace correlation are somewhat representing the slight improvement observed in our outcome.





\subsection{A note about generalization}
\label{sec:experiments-generalization}

In real applications, filtering ground roll is not done individually for each gather but using a very complex and tuned processing batch, that takes a lot of time, but once finished is applied to all gathers at once. For ground roll filtering, creating a new training data for each survey is already solving the problem, and therefore generalization of trained models is an important feature for such filtering methods. 

\begin{figure}[!ht]
\centering
\includegraphics[width=0.7\linewidth]{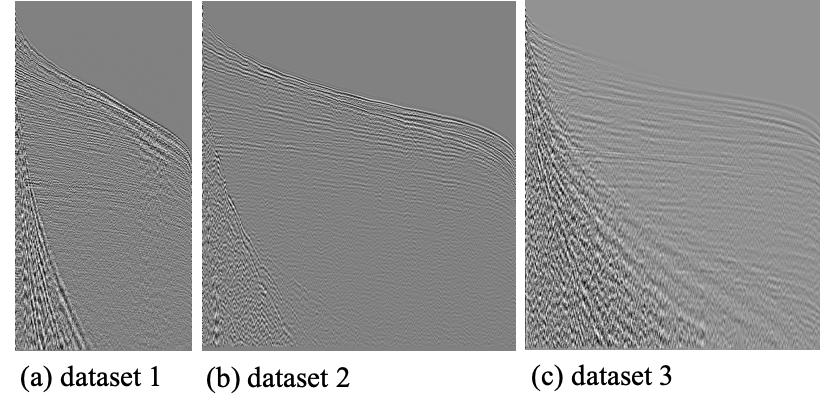}
\caption{Examples of shot gathers from different datasets: (a) dataset 1: used for training the model, (b) dataset 2: similar geology from dataset 1 in different area, (c) dataset 3: different geology in different area.}
\label{fig:fine_det}
\end{figure}

With this in mind, we considered two other datasets to evaluate the generalization of our trained model. One with similar geology, but at a different location, and the other with a different geology. The results presented in table \ref{fig:results-tab} show that for similar geologies our model behaves very well (datasets 1 and 2), but as expected, in different geologies the results are poor, specially the power spectrum frequency analysis (dataset 3). This indicates that for real applications one would need a few trained models to cope with geologies from a given region of interest, but in cases of similar geologies reasonable and useful results are expected from the trained models.

\begin{table}[ht]
\centering
\caption{Average and standard deviation of evaluation metric scores for power spectrum, amplitude and trace correlation with reference. Results were computed using the same model trained from dataset 1.}
\begin{tabular}{|c|c|c|c|}
\hline
Dataset & Pwr Spectrum & Amplitude & Correlation \\ 
\hline
1-HMAO & 118.99\% (35.65) & 104.51\% (12.63) & 94.86\% (1.96)  \\ 
2-HMAO & 98.15\% (28.94) & 92.57\% (22.24) & 93.87\% (2.58)  \\ 
3-YaNAO & 8.32\% (20.58) & 85.59\% (21.52) & 69.73\% (5.84)  \\ 
\hline
\end{tabular}
\label{fig:results-tab}
\end{table}

\section{Conclusions}
\label{sec:conclusions}

In this paper we introduced an approach for ground roll detection and suppression supported by different convolutional neural networks. Our results indicate the potential of the proposed pipeline in comparison with human expert filtering. We also briefly discussed generalization and presented good results in similar geologies. 

Further research includes a broader evaluation of models generalization and the influence in performance of sensors, different geological structures and geolocation of training data. Proposing and evaluating fully unsupervised methods for ground roll suppression seems to be also promising.

\addtolength{\textheight}{-12cm}   









\end{document}